\documentclass[twocolumn,showpacs,preprintnumbers,amsmath,amssymb,prl]{revtex4}
\usepackage{graphicx}% Include figure files
\usepackage{dcolumn}% Align table columns on decimal point
\usepackage{bm}% bold math

\begin{document}

\preprint{APS/123-QED}

\title{Nonequilibrium Casimir-Polder Force in Non-Stationary Systems}
% Force line breaks with \\

\author{Ryan O. Behunin$^1$ and B. L. Hu$^{1,2}$}

\affiliation{$^1$ Maryland Center for Fundamental Physics and $^2$ Joint Quantum Institute, \\
University of Maryland, College Park, Maryland, 20742}
\date{July 31, 2009}

%\begin{abstract}
%Employing the influence functional formalism we derive the Casimir-Polder (CP) force, the force felt by an atom near a substrate, for an ideal conductor. This fully nonequilibrium formulation provides an equation of motion for the atom's mean position. We compute the CP force for zero and finite temperature. By treating the center of mass motion as a dynamical variable we compute a Langevin equation to describe the stochastic fluctuations of the atom's trajectory about its semi-classical value which could be relevant to experiments with cold atomic gases.
%\end{abstract}
\begin{abstract}
Recently the Casmir-Polder force felt by an atom near a substrate under nonequilibrium stationary conditions has been studied theoretically with macroscopic quantum electrodyanamics (MQED) and verified experimentally with cold atoms. We give a quantum field theory derivation of the Langevin equation describing the atom's motion based on the influence functional method valid for fully nonequilibrium (nonstationary) conditions.  
The noise associated with the quantum field derived from first principles is generally colored and nonlocal,
%The noise associated with the quantum field is generally colored and nonlocal, 
which is at variance with the `local source hypothesis' of MQED's generalization to nonequilibrium conditions.
%The Langevin equation predicts the dispersion in the distribution of atoms in the long time limit. 
Precision measurements on the shape deformation of an atomic gas as a function of its distance from a mirror would provide a direct check of our predictions based on this Langevin equation.
%For the special case of trapped atoms we recover the CP force valid for nonequilibrium and nonstationary setups. We also derive the Langevin equation describing the lowest order fluctuations in the atom's trajectory.

\end{abstract}
\maketitle

%%%%%%%%%%%%%%%%%%%%%%%%%%%%%
The Casimir-Polder (CP) force  \cite{CasPol} between a neutral atom and a mirror or a dielectric surface \cite{Lifshitz} has drawn renewed attention of theorists \cite{APS05,APSS08,Buh} because of real possibilities of detection \cite{Cornell}.
%such as the Casimir \cite{Casimir} force between two mirrors in a vacuum, the dynamical Casimir (DC) effect \cite{DCE}  when one of them is moving, and the Casimir-Polder (CP) \cite{ShresHu} dispersive force between an  atom and magneto-electric surfaces.
At short distances ($<$100 nm) CP-Lifshitz type forces dominate the interaction between neutral bodies making them a relevant or even essential factor in the design of micromechanical devices, traps for cold atoms and in precision measurements for the detection of deviations from known forces.

At a finite temperature this force has two components originating from the vacuum  and thermal fluctuations of the electromagnetic field, the latter is known as the Lifshitz force.  Two major theoretical approaches are used: quantum field theory (QFT) under external conditions \cite{BorMos}, which starts from microphysics based on QFT,  and macroscopic quantum electrodynamics (MQED) \cite{Lifshitz}, which is in the form of a linear response theory with an added stochastic source. MQED has been skillfully applied  \cite{APS05,APSS08,Buh} to the situation of a substrate at a different temperature from the field under nonequilibrium stationary conditions.  However one key assumption of MQED, that the fluctuations are local, has not been justified.  It remains a challenge to bridge these two approaches, to remove or justify such an assumption, and to generalize to fully nonequilibrium conditions for non-stationary systems. This is the aim of our research program. 

In this letter we lay out the basic structure of such a theory based on quantum open systems conceptual framework and the influence functional (IF)  formalism \cite{FeyVer}. Our nonequilibrium formulation recovers the well-known CP force on  an atom when the trajectory of the atom is stationary [Eq. (\ref{NFCP},\ref{FFCP})], and the thermal CP force in a finite temperature field [Eq. (\ref{FCPb1},\ref{FCPb2})].  More importantly, it gives a first-principles derivation of a Langevin equation which describes the atom's stochastic motion. The noise in this equation is generally colored and nonlocal, at variance with the main assumption of MQED. Our result for the dispersion of an atomic cloud could motivate experiments designed to measure its change in shape as a function of its distance from the mirror.

{\it The atom and its trajectory in a quantum field with boundary}
% We consider a tripartite system consisting of the photon field (environment), $A_\mu$, coupled to a 3d harmonic oscillator,  $\vec{Q}$, in the dipole approximation with center of mass motion $\vec{z}$.
We model the internal degrees of freedom (idf) $\vec{Q}$ of an atom by a (3-dimensional) harmonic oscillator with natural frequency $\Omega$. 
%\footnote{For the exploration of these effects we shall thus use the word atom and oscillator interchangeably.}. 
The atom moves on a trajectory $\vec{z}(t)$ in a quantum  field $A^\mu$, the electromagnetic vector potential, in the presence of a mirror. The dynamics of the system is determined self-consistently by allowing these three variables ($\vec{Q},A^\mu, \vec{z}$) to negotiate amongst themselves. Even for a stationary atom it is necessary in the set up of the problem to assume its position $\vec{z}$ to be a dynamical variable so its resultant trajectory comes from the  mutual interactions with the other two variables.
The action describing the entire system is
$S[Q, A^\mu,\vec{z} ]=S_Q[\vec{Q}]+S_{E}[A^\mu]+S_Z[\vec{z}]+S_{int}[Q,A^\mu,\vec{z}]$
(subscript $E$ stands for the electromagnetic field which serves as an environment) with the action for the oscillator given by
$S_Q[\vec{Q}]=\frac{m}{2}\int d\lambda[\dot{\vec{Q}}(\lambda)^2-\Omega^2\vec{Q}(\lambda)^2]$
where $m$ is the oscillator's reduced mass and $\lambda$ parameterizes its worldline. The photon field action is given by
$S_{E}[A^\mu]=-\frac{1}{4}\int d^4x F_{\mu\nu} F^{\mu\nu}$
where  $F_{\mu \nu}=\partial_{\mu} A_{\nu}-\partial_{\nu} A_{\mu}$ is the field strength tensor. The action for the motion of the atom's center of mass $M$ is
$
S_Z[\vec{z}]=\int d\lambda [ \frac{1}{2}M {\dot{\vec{z}}}^2(\lambda)-V[\vec{z}]  ]
$
where $V[\vec{z}]$ is an external potential.

In the dipole approximation, the Hamiltonian for an atom interacting with the photon field takes the form $-\vec{d}\cdot \vec{E}$ where $\vec{d}$ is the dipole moment of the atom and $\vec{E}$ is the electric field. In this spirit we define the interaction action
$
S_{int}[\vec{Q}, \vec{z} , A_\mu]= q\int d\lambda {Q}^k(\lambda) E_k[z^{\mu}(\lambda)]
$ where Greek indices denote spacetime components, $0$ for time, Roman indices will be reserved for purely spatial components, and the Einstein summation convention is used throughout.
%%%%%%%%%%%%%%%%%%%%%%%%%%%%%%%

$\it{World \ Line \ Influence \ Functional}$
Assume that at time $t_{in}$ the quantum statistical state of the oscillator, trajectory and field is described by a density operator $\hat{\rho}(t_{in})$. This state is unitarily evolved from the initial time $t_{in}$ to a later time $t_f>t_{in}$, and can be expressed in terms of path integrals by considering matrix elements in an appropriate basis.
The overall influence of the field on the dynamics of the atom is obtained by coarse-graining over the field variables resulting in the field-reduced density matrix \cite{HPZ1},
$
\rho_r(\vec{Q}_f,\vec{Q}'_f;\vec{z}_f,\vec{z}'_f;t_f)=\int d\vec{Q}_{in}   d\vec{Q}'_{in} \int d\vec{z}_{in}   d\vec{z}'_{in}  \int_{\vec{Q}_{in}}^{\vec{Q}_f}\mathcal{D}\vec{Q}  \int_{\vec{Q}_{in}'}^{\vec{Q}'_f} \mathcal{D}\vec{Q}' \int_{\vec{z}_{in},\vec{z}_{in}'}^{\vec{z}_f,\vec{z}'_f}\mathcal{D}\vec{z} \mathcal{D}\vec{z}'  \\  \times e^{i(S_Q[\vec{Q}]+S_Z[\vec{z}]-S_Q[\vec{Q}']-S_Z[\vec{z}'])} \rho_Q(\vec{Q}_{in},\vec{Q}_{in}';t_{in}) \\ \times \rho_Z(\vec{z}_{in},\vec{z}_{in}';t_{in}) \mathcal{F}[{J}^{\mu-}, {J}^{\nu+}]
$,
where $\mathcal{D}k$ is the measure for a path integral over the space of functions. This introduces the influence functional (IF) $\mathcal{F}[{J}^{\mu-}, {J}^{\nu+}]$ \cite{FeyVer}.

For the coupling given above and %(\ref{intaction}),
assuming an initially uncorrelated and Gaussian state the influence functional can be calculated exactly and is given by
\begin{eqnarray}
\label{IF}
\mathcal{F}[{J}^{\mu-}, {J}^{\nu+}]=\exp\bigg\{i\int d^4y \ J^{\mu-}(y) \ \ \ \ \ \ \  \nonumber \\ \times  \int d^4y' [D^{ret}_{\mu \nu}(y,y')J^{\nu+}(y')  +\frac{i}{4}D^H_{\mu \nu}(y,y')J^{\nu-}(y')]\bigg\}.
\end{eqnarray}
Here the current density is $J_\mu(x)=-q\int d\lambda( \partial_0 \eta_{j \mu}+\partial_j \eta_{0 \mu} )\delta^4(x^\mu-z^\mu(\lambda))Q^j(\lambda)$,  $J^+=(J+J')/2$ and $J^-=J-J'$ are its difference and semi-sum, respectively, and $\eta_{\mu \nu}= diag(-1,1,1,1)$ is the metric for Minkowski space. $D^{ret}_{\mu \nu}(y,y')$ and $D^H_{\mu \nu}(y,y')$ are the retarded Green's function and Hadamard function for the field, repectively. They can be expressed in the Feynman gauge in terms of the corresponding Green's function for a massless scalar field in Minkowski space as $D^{ret}_{\mu \nu}(x,x')=\eta_{\mu \nu}G_{ret}(x,x')$ and $D^{H}_{\mu \nu}(x,x')=\eta_{\mu \nu}G_{H}(x,x')$. % \cite{BD}.

To find the combined influence that the oscillator and the field have on
the trajectory we need to further coarse grain the oscillator  degrees
of freedom resulting in the oscillator-reduced
influence functional (ORIF), $\mathcal{F}_Z[\vec{z}^-,\vec{z}^+]$.
\begin{eqnarray}
\label{orIF1}
\mathcal{F}_Z[\vec{z}^-,\vec{z}^+]=\int d\vec{Q}_f d\vec{Q}_{in}   d\vec{Q}'_{in} \int_{\vec{Q}_{in},\vec{Q}_{in}'}^{\vec{Q}_f,\vec{Q}_f}\mathcal{D}\vec{Q} \mathcal{D}\vec{Q}'  \nonumber \\ \times e^{i(S_Q[\vec{Q}]-S_Q[\vec{Q}'])} \rho_Q(\vec{Q}_{in},\vec{Q}_{in}';t_{in}) \mathcal{F}[J^{\mu-},J^{\nu+}]
\end{eqnarray}

Noting that we cannot trace over the oscillator variables in
(\ref{orIF1}) explicitly for arbitrary field boundary conditions, such as in the presence of a mirror, we proceed via a perturbative expansion in powers of the coupling.
Writing (\ref{orIF1}) in a more suggestive form $\mathcal{F}_Z[\vec{z}^{+},\vec{z}^{\mu-}]=\exp\{iS_{inf}[{z}^{\mu+},z^{\mu-};-i\frac{\delta}{\delta j^+_k},-i\frac{\delta}{\delta j^-_l}]\}f_o[\vec{j}^+,\vec{j}^-] |_{j^\pm=0}$, which defines the influence action, $S_{inf}[{z}^{\mu+},z^{\nu-}; Q^-_j, Q^+_k]=-i\ln \mathcal{F}[J^{\mu+},J^{\nu-}]$, and the IF for a three dimensional harmonic oscillator, $f_o[\vec{j}^+,\vec{j}^-]$. To factor the exponent out of the path integral $[Q^{k \pm}(\lambda)]^n$ is replaced with $\left(-i\frac{\delta}{\delta j^{\mp}_k(\lambda)} \right)^n f_o[\vec{j}^+,\vec{j}^-] |_{j^{\pm}=0}$.
For a Gaussian initial state $f_o[\vec{j}^+,\vec{j}^-]$ can be evaluated exactly $f_o[\vec{j}^+,\vec{j}^-] =\mathcal{N} \exp \{ i\int d\lambda d\lambda' [ \vec{j^-}(\lambda)\cdot\vec{j^+}(\lambda')g_{ret}(\lambda,\lambda')+\frac{i}{4}\vec{j^-}(\lambda) \cdot\vec{j^-}(\lambda')g_{H}(\lambda,\lambda')] \}$
where $g_{ret}(\lambda,\lambda')$ and $g_{H}(\lambda,\lambda')$ are the retarded and Hadamard Green's functions for a one dimensional harmonic oscillator with natural frequency $\Omega$, $\mathcal{N}$ is a normalization constant, and the dot product is taken with respect to a 3 dimensional Euclidean metric.

Expanding (\ref{orIF1}) to lowest order in the coupling and partially resumming gives the ORIF to second order in the coupling $-i \ln \mathcal{F}_Z[\vec{z}^+,\vec{z}^-]  \approx    S_{inf}[z^{\mu+},z^{\mu-};-i\frac{\delta}{\delta j^+_k},-i\frac{\delta}{\delta j^-_l}]f_o[\vec{j}^+,\vec{j}^-] |_{j^\pm=0}   $ allowing us to write the reduced density matrix describing the center of mass motion.
\begin{eqnarray}
\label{rhor}
\rho_r(\vec{z}^+_f,\vec{z}^-_f;t_f)=\int d\vec{z}^+_{in}   d\vec{z}^-_{in}  \int_{\vec{z}^+_{in},\vec{z}^-_{in}}^{\vec{z}^+_f,\vec{z}^-_f}\mathcal{D}\vec{z}^+ \mathcal{D}\vec{z}^-    \nonumber \\ \times e^{i(S_Z[\vec{z}]-S_Z[\vec{z}'])} \rho_Z(\vec{z}^+_{in},\vec{z}^-_{in};t_{in}) \mathcal{F}_Z[\vec{z}^+,\vec{z}^-] \ \ \ \ \ \ \ \ \ \ \
\end{eqnarray}

%%%%%%%%%%%%%%%%%%%%%%%%%%%
{\it Atom's mean trajectory}
%%%%%%%%%%%%%%%%%%%%%%%%%%%
The complex norm of the ORIF, $|\rho_r|\propto \exp\{- \int d\lambda d\lambda' z^{k-}(\lambda) N_{kj}(\lambda,\lambda') z^{j-}(\lambda') \}$ is non-vanishing and strongly suppressed for large values of the off diagonal elements, $\vec{z}^-=\vec{z}-\vec{z}'$, as is indicative of decoherence of the quantum trajectory. $N_{kj}$ is a symmetric positive definite kernel quantifying the noise in the oscillator and field.

Decoherence of the system due to its interactions with the quantum
fluctuations of the environment and oscillator permits the existence
of a semi-classical limit for the oscillator's path through space.
Using a saddle-point approximation to evaluate (\ref{rhor}) about its
classical solution, $z^k_{cl}(\lambda)\equiv \bar{z}^k$, one can show
that the semi-classical dynamics is determined from the variation
${\delta S_{CGEA}[z^{k+},z^{k-}]}/{\delta z^{j-}(\tau)}|_{z^{k-}=0}=0$
where the so-called coarse grained effective action is given by
$S_{CGEA}[z^{k+},z^{k-}]=S_Z[\vec{z}]-S_Z[\vec{z}']-i\ln \mathcal{F}_Z[\vec{z}^+,\vec{z}^-]$.

%Suppression of the reduced density matrix by large values of $\vec{z}^-$ justifies expanding $S_{CGEA}$ for small $\vec{z}^-$.
%To capture the lowest order information about fluctuations of the semiclassical solution we expand the $S_{CGEA}$ to second order in $\vec{z}^-$.
Varying $S_{CGEA}$ with respect to $\vec{z}^-$ we obtain the mean (semi-classical) equation of motion \cite{JH1}
\begin{equation}
\label{sceom}
M\ddot{z}_k (\tau)+\partial_k V[\vec{z}(\tau)]=f_k(\tau)
\end{equation}
where the effective force, $f_k(\tau)$ (including back-action effects), has the form
\begin{eqnarray}
\label{IFo}
{f}_k(\tau)=\frac{q^2}{2} \int_{\lambda_{in}}^{\lambda_f} d\lambda \int_{\lambda_{in}}^{\lambda_f} d\lambda' \delta^{ij} \delta(\lambda-\tau) \partial_k \kappa ^\alpha_{i}\kappa ^\beta_{j'}  \ \ \ \ \ \nonumber \\   \bigg\{ g_H(\lambda,\lambda')  D^{ret}_{\alpha\beta}(z^\alpha(\lambda),z^\alpha(\lambda')) \nonumber \\ + g_{ret}(\lambda,\lambda') D^{H}_{\alpha\beta} (z^\alpha(\lambda),z^\alpha(\lambda')) \bigg\}
\end{eqnarray}
where $\kappa ^\mu_j=\partial_0 \eta^\mu_j+\partial_j \eta^\mu_0$. Take caution to evaluate the derivatives before the particle trajectory is placed into the various kernels.

The influence or back-action force on the oscillator trajectory describes dissipation and radiation reaction as
well as the forces due to constraints on the field. The first two
effects must be taken into account when atom motion comes into play.
In the following we assume an appropriate form for $V[\vec{z}]$ so that (\ref{sceom}) admits
static solutions where dissipative effects may be ignored.

%%%%%%%%%%%%%%%%%%%%%%%%%%%%%%%
$\it{Casimir-Polder \ Force}$
The placement of a mirror in the $z=0$ plane constrains the transverse components of the electric field to
vanish there,  and will lead to forces on the atom. This boundary
condition can be accommodated by appealing to the method of images.
Thus, a dipole near a mirror will be attracted to its
image on the other side, a classical electrostatic treatment for a permanent dipole
gives a $1/z^4$
dependence. When finite light propagation time and  quantum
fluctuations are accounted for this attractive force takes a modified
form, $1/z^5$, in the far field limit where the distance from the
mirror is much greater than the
period of the oscillator (c=1). This is the Casimir-Polder force.

From the Green's function point of view the field constraint can be satisfied
by pairing every Green's function with an image term i.e.
$G(\sigma)\rightarrow G(\sigma)-G(\tilde{\sigma})$ where $\sigma(x,x')$
is Synge's worldfunction defined to be half the geodesic distance between
$x$ and $x'$ and $\tilde{\sigma}(x,x')=\sigma(x,x')+2zz'$. The new terms, $F^{CP}_k$, due to the
presence of a mirror are responsible for the CP effect.
\begin{eqnarray}
\label{InfFo}
F^{CP}_k(\tau)=\frac{q^2}{2} \int_{0}^{\tau-\lambda_i} ds \ \delta^{ij} \partial_k \kappa ^\alpha_{i}\kappa ^\beta_{j'} \nonumber \\  \bigg\{ g_H(s) \tilde{D}^{ret}_{\alpha\beta}(\tilde{\sigma}[z^\alpha(\tau),z^\alpha(\tau-s)]) \nonumber \\  + g_{ret}(s)  \tilde{D}^{H}_{\alpha\beta} (\tilde{\sigma}[z^\alpha(\tau),z^\alpha(\tau-s)]) \bigg\}
\end{eqnarray}
where a prime denotes differentiation with respect to the second argument. To accommodate the boundary conditions on the field the tensor structure of the image term changes $\tilde{D}_{\alpha\beta}(\tilde{\sigma})=-(\eta_{\alpha \beta}-2\hat{z}_\alpha \hat{z}_\beta)G(\tilde{\sigma})$
where $\hat{z}_\alpha=(0,0,0,1)$.

The effective force from the image takes on the general form
\begin{eqnarray}
\label{CPF1}
F^{CP}_{k}(\tau)=\frac{q^2}{2 m\Omega}\eta^{\mu \nu}  \int_{0}^{\tau-\lambda_i} ds \bigg[ \tilde{\sigma}_k \tilde{\sigma}_\mu \tilde{\sigma}_{\nu'} \bigg(\frac{d}{d\tilde{\sigma}}\bigg)^3 \nonumber \\ +(\tilde{\sigma}_{\mu \nu'}\tilde{\sigma}_k+\tilde{\sigma}_{\mu k}\tilde{\sigma}_{\nu'}+\tilde{\sigma}_{k\nu'}\tilde{\sigma}_\mu ) \bigg(\frac{d}{d\tilde{\sigma}}\bigg)^2  \bigg] \nonumber \\  \bigg[ \cos \Omega s  \ G_{ret}(\tilde{\sigma})  + \sin \Omega s  \ G_H(\tilde{\sigma}) \bigg] \ \ \ \ \ \ \
\end{eqnarray}
where $d/d\tilde{\sigma}$ operates only on the Green's functions for the field and $\tilde{\sigma}_k=\partial_k \tilde{\sigma}$. To find an explicit expression for the Casimir-Polder force we evaluate (\ref{CPF1}) for a static trajectory, $z^\mu(\tau)=(\tau,\vec{z})$ and $\dot{z}^\mu(\tau)=(1,\vec{0})$. We find an analytic expression for the CP force in the long time limit when the field has dressed the atomic ground state and the field and oscillator were initially in the their respective ground states. The CP-force has two contributions $F^{CP}_{k}(\tau)=F^{CP1}_{k}(\tau)+F^{CP2}_{k}(\tau)$.
\begin{eqnarray}
\label{CPF1s}
F^{CP1}_{k}(\tau)=\frac{q^2 \eta_{kz}  z}{2^7 \pi m\Omega}  \ \theta(\tau-2z) \bigg\{ z^2, \bigg(\frac{1}{ z}\frac{d}{d z}\bigg)^3 \bigg\} \frac{\cos 2\Omega z}{z}
\end{eqnarray}
\begin{eqnarray}
F^{CP2}_{k}=-\frac{ q^2}{32 \pi^2 m \Omega z^5} \eta_{k z}  \
\bigg[ 8\Omega z+6(1-2\Omega^2z^2)f(2\Omega z) \nonumber \\ -4\Omega z(2 \Omega^2 z^2 -3)g(2\Omega z) \bigg] \ \ \ \
\end{eqnarray}
where $f(x)(g(x))$ is the auxiliary function for the cosine (sine) integral function, %\cite{AS},
and $\{A,B\}$ is the anticommutator of $A$ and $B$.

%The interpretation for each component can be taken from the kernels contained in each.
Here $F^{CP1}$ is derived from the term containing the retarded Green's function for the field and so is responsible for the electrostatic contribution to the CP force and the near field behavior. $F^{CP2}$ is the dispersive part of the force because it contains the field Hadamard function.
%, and is analogous to the expression derived using energy gradient methods. Indeed $F^{CP2}$ agrees with the exact analytic form of the the CP-force expression derived using energy gradient methods for a two-level atom.
For interactions linear in the oscillator coordinate as assumed here,  in perturbation theory the quantum amplitude to go from the ground state to any but the first excited state vanishes.
%So by calculating the force perturbatively we have restricted to the lowest two energy levels of the oscillator.
Thus the agreement of the present HO results with previous results for two level atoms using energy gradient methods \cite{Passant} is not surprising.

In the near and far field limits we recover the asymptotic expressions
\begin{eqnarray}
\label{NFCP}
\Omega z <<1 \ \ \ \   F^{CP}_z  \approx  -\frac{3 q^2}{32 \pi m \Omega z^4} 
\\
\label{FFCP}
\Omega z >>1 \ \ \ \   F^{CP}_z  \approx  -\frac{3 q^2}{8 \pi^2 m \Omega^2 z^5}
\end{eqnarray}
where our results agree with the literature if we identify the static polarizability, $\alpha$,  with $q^2/4\pi m\Omega^2$, this form for $\alpha$ can be argued by examining the static solutions to the classical equations of motion.

%%%%%%%%%%%%%%%%%%%%%%%%%%%%%%%
$\it{Thermal \ CP \ force}$
%A thermal CP force arises when the atom/oscillator is in a thermal field with a mirror present. This is easily %obtained from our fully dynamical or nonequilibrium formulation established above.
The form of the CP force in a thermal field can be taken directly from (\ref{InfFo}) with all Green's functions replaced with their appropriate finite temperature version.
%Here the power of our method to analyze full nonequilibrium conditions manifests. In our approach we time evolve an
The assumption of an initially factorized density matrix allows us to independently choose the initial oscillator and field state. Choosing the oscillator and field to be in thermal states of different temperature (with inverse temperatures $\bar{\beta}, \beta$ respectively) gives rise to two distinct thermal contributions to the CP force.

The retarded Green's functions appearing in (\ref{InfFo}) will not contribute to the thermal effects as they are state independent. Modifications due to an initially thermal state will arise from the Hadamard functions only.
%Focusing here we can isolate the changes in the physics when we treat the field at finite temperature.
The thermal Hadamard function for the field can be found by imposing a periodicity condition on the imaginary time \cite{BD}.  For a harmonic oscillator it can be calculated directly $g^{\bar{\beta}}_H(\tau,\tau')= \coth ( \bar{\beta} \Omega/2 ) \cos \Omega (\tau-\tau')/m\Omega$
%From the form of $g^{\bar{\beta}}_H$ we can immediately write down the thermal behavior of $F^{CP1}_k$.
%\footnote{The factor of $\coth (\bar{\beta} \Omega /2)$ is independent of any integration variables and so simply scales the zero-temperature result.}.
This gives
\begin{equation}
\label{FCPb1}
F^{CP\bar{\beta}1}_k= \coth (\bar{\beta} \Omega /2) F^{CP1}_k 
\end{equation}
\begin{eqnarray}
\label{FCPb2}
F^{CP\beta2}_k(\tau)=\frac{q^2}{2 m \Omega}  \sum_{k=-\infty}^{\infty} P.V. \int_{0}^{\tau-\lambda_i} ds \nonumber \\ \times \sin \Omega s \ \partial_k\partial^\nu \partial_{\nu'} G_H(t+ik\beta(\vec{z}),\vec{z},\tilde{z}')
\end{eqnarray}
where $\tilde{z}=(t,x,y,-z)$. We have included the generalized case of a field state of spatially nonuniform temperature i.e. $\beta \rightarrow \beta(\vec{x})$ as it is is Gaussian in field variables.

In the high temperature, long time and far field limit we arrive at
\begin{equation}
\label{ }
F^{CP\beta2}_k  \approx - \frac{3q^2\eta_{kz}}{16 \pi \beta m \Omega^2} \frac{1}{z^4} =-\frac{3}{4}\eta_{k z} \frac{\alpha}{\beta z^4}.
\end{equation}

\begin{figure}
\begin{center}
\includegraphics[width=2.5in]{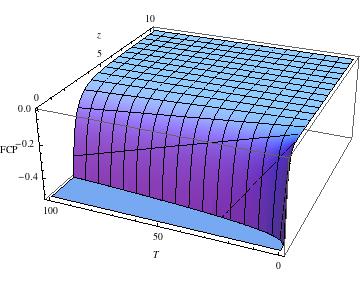}

\caption{
 Plot of thermal C-P force in units of $\hbar q^2/m$, for  $\Omega = 1$, in the long time limit against perpendicular distance z (in units $c/\Omega$) of the atom from a plane mirror and absolute temperature T (in $  (\hbar \Omega/k_B)K $)}
\label{ }
\end{center}
\end{figure}

%%%%%%%%%%%%%%%%%%%%%%%%%%%%%%%%
$\it{Stochastic \ trajectory}$
%When the coarse-grained oscillator and quantum field fluctuations induce small perturbations about the semi-classical solution $\bar{z}_k(\tau)$ the trajectory can be shown to evolve within a stochastic semi-classical limit. In such a case, these fluctuations manifest as classical stochastic noise in the oscillator's path through space.
The IF can produce a Langevin equation for the trajectory with deviations from the mean caused by the quantum field fluctuations.  It is given by
\begin{equation}
\label{le}
M\ddot{\tilde{z}}_k+\partial_\alpha \partial_k V[ \bar{z}_i] \tilde{z}^\alpha-   \partial_\alpha f_k[\bar{z}_i] \tilde{z}^\alpha = \xi_k [\bar{z}_i].
\end{equation}
%The distribution of the stochastic source is given by the see (ref), and
The key link in identifying a classical stochastic source (noise) from a quantum field is provided by the Feynman-Vernon identity for Gaussian integrals \cite{FeyVer}. The two-point function for this classical stochastic source is related to the noise kernel by
\begin{eqnarray}
\label{ }
\left<\{ \xi_k[z^\alpha(\lambda)], \xi_j[z^\alpha(\lambda')]\} \right> = \frac{q^2}{2}\delta^{mn} g_H(\lambda,\lambda') \nonumber \\ \times  \partial_{k}  \partial_{j'} \kappa ^\alpha_{m}\kappa ^{\beta'}_{n}D^H_{\alpha\beta}[z^\alpha(\lambda),z^\alpha(\lambda')].
\end{eqnarray}

The Langevin equation enables us to calculate the dispersion of the atom's trajectory,  $\left<\Delta\vec{z}^2(\tau) \right>$, which is defined as the effective distance from the mean value that an ensemble of stochastic realizations takes. As the noise kernel contains the Hadamard function for the field it is sensitive to the boundary condition at $z=0$. The image term present will make the distribution of noise vary with the distance from the mirror and in turn the dispersion in the atoms' positions as well. This manifests
% The correction to the dispersion in the presence of a mirror can be calculated directly and is proportional to the
as a fractional change in volume of a gas of noninteracting atoms. If we trap the atoms in a harmonic potential with frequency $\Omega_k$ in the kth direction, such that $|\Omega_k^2-\Omega^2|>>q^2/m\Omega^3 M z^6$ then the dissipation can be ignored in the final expression for the dispersion and we can directly compute the far-field long-time limit. The dispersion in the z-direction is given by
\begin{equation}
\label{dispz}
\left<\delta \tilde{z}^2\right>_\xi\approx -\frac{15 q^2}{16 \pi^2 m \Omega M^2}\frac{1}{(\tilde{\Omega}_z^2-\Omega^2)^2}\frac{1}{z^6}
\end{equation}
where $\tilde{\Omega}_z$ is the trapping potential frequency in the presence of a mirror. The parallel components can be obtained from (\ref{dispz}) by dividing by $-15$ and substituting the trap potential frequency for the unperturbed dimension. The expression for the dispersion shows that the presence of the mirror leads to a focusing in the perpendicular direction and a broadening in the parallel directions.
%Recently reported techniques for high precision measurements of the position of a single atom \cite{Orozco} lends hope to measurements with increased accuracy on the dispersion of an ensemble of atoms.

In conclusion we have derived from first principles the semiclassical and stochastic equations for an atom's motion near a mirror under fully nonequilibrium conditions. As a quantum field theory derivation of the stochastic source is given there is no need and no room for a `local source hypothesis' which generalizes MQED to nonequilibrium conditions.
%For the special case of trapped atoms we recover the CP force valid for nonequilibrium and nonstationary setups. We also derive the Langevin equation describing the lowest order fluctuations in the atom's trajectory.  
Precision measurements (see e.g., \cite{Orozco}) in the shape deformation of an atomic gas near a mirror as a function of atom-mirror spacings would provide a direct check against our theoretical predictions.
 
%\begin{acknowledgements}
We wish to thank Chad Galley for useful discussions. This work is supported in part by NSF Grants No. PHY-0426696, 0601550, 0801368, and MCFP.
%\end{acknowledgements}
%%%%%%%%%%%%%%%%%%%%%%%%%%%%%%%
%\section{Conclusion} We have derived the CP effect by employing the influence functional formalism.	% (uses file "plain.bst")

\end{document}